\documentclass[preprint,showkeys,showpacs,preprintnumbers,amsmath,amssymb,prl]{revtex4}

\usepackage{graphicx}
\usepackage{dcolumn}
\usepackage{bm}
\usepackage{amsmath}
\usepackage{color} 

\def\a{\alpha}
\def\t{\tau}
\def\l{\lambda}
\def\k{K}
\def\d{\delta}
\def\e{\epsilon}
\def\g{\gamma}

\def\Jij{J_{ij}}
\def\eij{\e_{ij}}

\begin{document}

\preprint{?APS/123-QED?}

\title{Critical Avalanches and Subsampling in Map-based Neural Networks}

\author{M. Girardi-Schappo}
\affiliation{
Departamento de F\'isica, Universidade Federal de Santa Catarina, 88040-900, Florian\'opolis, Santa Catarina, Brazil
}

\author{O. Kinouchi}
\affiliation{
Departamento de F\'isica, FFCLRP, Universidade de S\~ao Paulo, 14040-900, Ribeir\~ao Preto, S\~ao Paulo, Brazil
}

\author{M. H. R. Tragtenberg}
\email{marcelotragtenberg@gmail.com}
\affiliation{
Departamento de F\'isica, Universidade Federal de Santa Catarina, 88040-900, Florian\'opolis, Santa Catarina, Brazil
}

\date{\today}

\begin{abstract}
We investigate the synaptic noise as a novel mechanism for creating critical avalanches in the activity of neural networks.
We model neurons and chemical synapses by dynamical maps with a uniform noise term in the synaptic coupling. 
An advantage of utilizing maps is that
the dynamical properties (action potential profile, excitability properties, post synaptic potential summation etc.) 
are not imposed to the system, but occur naturally by solving the system equations. We discuss the relevant neuronal and synaptic properties to achieve the critical state. We verify that
networks of excitatory by rebound neurons with fast synapses present power law avalanches. 
We also discuss the measuring of neuronal avalanches by subsampling our data, 
shedding light on the experimental search for Self-Organized Criticality in neural networks.
\end{abstract}

\maketitle
The hypothesis of Self-Organized Critical (SOC) neural networks is based on theoretical considerations made in the 
90's \cite{linkenkaerRef141,stassiBrain,herzCML} and 
supported by experimental data obtained in the last decade \cite{chialvoReview,wernerFractais,plenzBenefits,beggsCritical}. In particular, the observation of neuronal avalanches motivated
the search for computational models presenting this phenomenon
\cite{beggsPlenz2003,kinouchiCopelli,socPlasticity,abbottCritico,levina,ribeiroCopelli}.
The key interest in these simulations is to find what are the conditions for the occurrence of power laws
in the size and duration distributions of avalanches.
Moreover, some authors showed that the critical state may optimize the dynamical (input) range
\cite{kinouchiCopelli,plenzDynRange}, the memory and learning processes \cite{socPlasticity}, 
and the computational power of the brain \cite{wernerFractais,plenzBenefits,beggsCritical}. However, up to now, the
computational models rely on very simplified neuron models like branching processes \cite{beggsPlenz2003},
cellular automata \cite{kinouchiCopelli,ribeiroCopelli} or integrate-and-fire neurons \cite{levina}.

\begin{figure}[b!]
	\begin{center}
	\includegraphics[width=50mm]{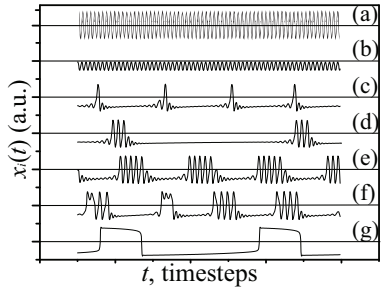}
	\end{center}
	\caption{\label{fg:ktzOverview}Examples of KTz behaviors (Eq. \ref{eq:ktzModel}) for $\k = 0.6$, $I_i(t)=0$. When not specified, $T=0.35$ and $\d=\l=0.001$.
	(a) fast spiking ($x_R=-0.2$,$T=0.45$); (b) subthreshold oscillations ($x_R=-0.5$,$T=0.45$); (c) slow spiking
	($x_R=-0.62$,$\d=\l=0.003$); (d) slow bursting ($x_R=-0.6$); (e) fast bursting ($x_R=-0.45$); (f) chaotic bursting
	($x_R=-0.4$,$T=0.322$); (g) cardiac-like spiking ($x_R=-0.5$,$T=0.25$). $x(t)$ is the membrane potential in arbitrary units.}
\end{figure}

Besides these simple approaches, neurons may be modeled by differential equations \cite{synapticModelSchutter}
or by discrete time maps \cite{modeloKT,ibarzMapas}.
Here, we use the KTz map \cite{modeloKTz2001,modeloKTz2004} which is a discrete time system with behavior similar to
the Hindmarsh-Rose 
model \cite{modeloHR}, a well accepted neuronal model of three ordinary differential
equations. KTz presents a very rich set of dynamical behaviors
(excitability, bursting, cardiac-like spikes, refractoriness, post-synaptic potential summation, etc.) with
a minimal set of parameters \cite{modeloKT,modeloKTz2001,modeloKTz2004}, see Fig. \ref{fg:ktzOverview}.
Maps are more efficiently solved by computers than differential
equations, as they have
discrete time dynamics \cite{ibarzMapas}.
The main advantage of choosing a complex model like KTz is that, unlike integrate-and-fire models, the neuronal-like dynamic properties 
are not artificially imposed to the system.

We connect the KTz neurons with a
Chemical Synapse Map (CSM) \cite{modeloKTz2001} in order to build a Coupled Map Lattice
\cite{cmlBook}.
Synaptic noise is present in every synaptic connection in the brain \cite{peretto}. Thus, we propose the
addition of noise in the synaptic coupling
as a novel mechanism for obtaining critical neuronal avalanches.

Concerning the experimental data for neuronal avalanches, we recall that it is
subsampled, since only a small fraction, $f$, of the neurons of the studied brain region is actually recorded. 
In such case, the statistical distributions generated by the
sampled neurons may not reproduce the distributions of the entire network activity.
Thus, we analyse the full and the subsampled data of our distributions of neuronal
avalanches with the same algorithm utilized to detect neuronal avalanches experimentally
\cite{violaSub,ribeiroCopelli}.

Each KTz neuron, labeled by an index $i=1,\cdots,N$, is given by the three-dimensional map
\begin{equation}
\label{eq:ktzModel}
\begin{array}[c]{l}
x_i(t+1)=\tanh\left[ \dfrac{x_i(t)- K y_i(t)+z_i(t)+v_i(t)}{T}  \right]\textnormal{,}\\
y_i(t+1)=x_i(t)\textnormal{,}\\
z_i(t+1)=(1-\d)z_i(t)-\l\left[x_i(t)-x_R\right]\textnormal{,}
\end{array}
\end{equation}
where
$x_i(t)$ represents the membrane potential of the $i$th neuron (fast
dynamics), $y_i(t)$
is the return variable and $z_i(t)$ is an adaptive variable (e.g. related to slow currents that governs the refractory period and bursting phenomena). 
The parameter $\delta$ is the inverse recovery time of $z(t)$, $K$ and $T$ are parameters of the fast subsystem
that define spiking, resting and spiking/resting coexistence regimes \cite{modeloKT}. The parameters 
$\lambda$ and $x_R$ control the slow spiking and bursting dynamics \cite{modeloKTz2001}.
All the currents received by the neuron,
whether synaptic currents or external stimuli, are summed up in
$v_i(t) = I_i^{ext}+\sum_{j}{I_{ij}^{syn}}$.

Chemical synaptic currents are modeled by \cite{modeloKTz2001}:
\begin{equation}
\label{eq:csmModel}
	\begin{array}[c]{l}
	I^{syn}_{ij}(t+1) = \left(1-\dfrac{1}{\t_1}\right) I^{syn}_{ij}(t) + h_{ij}(t)\textnormal{,}\\
	h_{ij}(t+1) = \left(1-\dfrac{1}{\t_2}\right)h_{ij}(t) + \Jij(t)\Theta(x_j(t))\textnormal{,}
	\end{array}
\end{equation}
where $I^{syn}_{ij}(t)$ is the synaptic current from neuron $j$ (presynaptic) to neuron $i$ (postsynaptic), $h_{ij}(t)$ is an
auxiliary variable for creating more complex synapses (e.g. double-exponential functions), $\t_1$ and $\t_2$ are time
constants for  $I_{ij}^{syn}$ and $h_{ij}$, $\Jij(t)$ is the coupling parameter
and $\Theta(x)$ is the step (Heaviside) function. Thus, if we start with $I^{syn} = h=0$, the $h$ variable is activated when
the membrane potential is depolarized above zero (which we define as an effective spike duration). This produces an activation
of the $I^{syn}$ current, which has a form of a discrete alpha function (for $\t_1=\t_2$) or a discrete double exponential
(for $\t_1 \neq \t_2$). Notice that the above equations
are not used to describe the time evolution of synaptic conductances (as usual) but the evolution of synaptic
currents, which is also an acceptable procedure in computational neuroscience \cite{synapticModelSchutter}.

Throughout this work, we call \textit{inhibitory} the synapses adjusted with parameter $J<0$, although one must bear in mind
that, in such case, the neurons are adjusted in an excitable by rebound regime. Thus, the synapses do not inhibit one cell's neighbors.
Instead, they may fire rebound spikes \cite{izhikevichDyn}.

\begin{figure}[t!]
	\begin{center}
	\includegraphics[width=60mm]{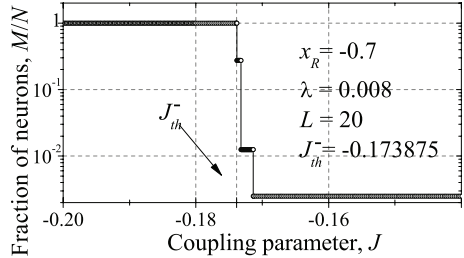}
	\end{center}
	\caption{\label{fg:phaseTrans}Fraction of neurons activated by a delta stimulus of intensity $I=0.1$ in a lattice
	with $L=20$ and neurons in regime I. $J_{th}^-=-0.173875$ is the threshold value below which the network is all activated and
	it has been determined computationally. It depends only on the neurons parameters.}
\end{figure}

In the homogeneous case, $\Jij(t)= J$, any network of excitable neurons with reciprocal synapses and free boundary conditions
presents a discontinuous bifurcation transition described
by the order parameter $M/N$ (the fraction of neurons that fired due to a delta stimulus, i.e. that participated in the
avalanche). We show in Fig. \ref{fg:phaseTrans} the case of
inhibitory synapses, in which there is a threshold $J=J_{th}^-<0$
that separates the state in which all the neurons take part
in the avalanche ($J<J_{th}^-$) from the state in which only the stimulated neuron, or a few neighbors, responds ($J>J_{th}^-$). A similar transition may occur for excitatory synapses
with $J=J_{th}^+>0$.

However, the homogeneous model cannot achieve a critical distribution of avalanches, because they are all of size $s=1$ or $N$ (disregarding
the small steps in the phase transition, which are independent of $N$).
Thus, motivated by the synaptic noise present in the brain, we propose an annealed coupling $\Jij(t)=J+\eij(t)$. In the case of inhibitory synapses, $J<0$ and
 $\eij(t)\in\left[R;0\right]$, since $J_{th}^-<0$. This models a uniform noise, different
for every connection $j\rightarrow i$ in the network, of maximal amplitude $|R|$, such that $|J+R|>|J_{th}^-|$. Then, the coupling fluctuates near $J_{th}^-$ in an uncorrelated manner, so we can define the probability that $|J_{ij}(t)|>|J_{th}^-|$:
\begin{equation}
\label{eq:excProb}
p = \dfrac{J+R-J_{th}^-}{R}\textnormal{.}
\end{equation}
The same holds for excitatory synapses ($J>0$), where $\eij(t)\in\left[0;R\right]$.
The synaptic parameters $J$ and $R$ are, in principle, our control parameters that are adjusted such
that there is a nonzero $p$. For convenience, we utilize $p$ instead of $R$ as control parameter.

\begin{figure}[b!]
	\includegraphics[width=85mm]{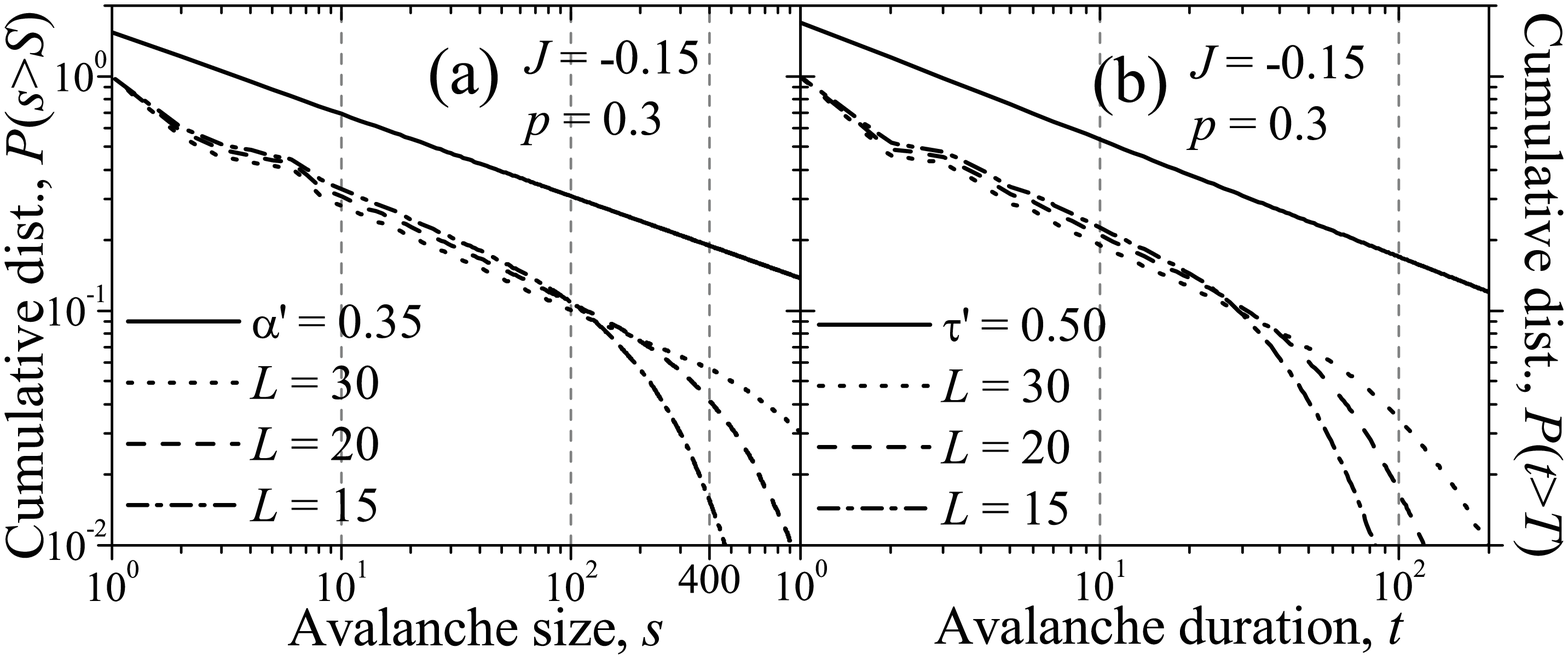}
	\caption{\label{fg:reg1L20L30SizeTime}(a) Avalanche size cumulative distributions and (b) avalanche duration cumulative distributions for neurons in regime I,
	$J=-0.15$, $p=0.3$ and $L=15$ (--$\,\cdot\,$--), $L=20$ (- - -) and $L=30$ ($\cdot\cdot\cdot$). Solid line is a power law fit. The exponent found for the avalanche sizes distribution is $\a=1.35$ and for the duration
	distribution is $\t=1.50$.}
\end{figure}

\textit{Results}. We plot the avalanche distributions as cumulative distribution
functions. This representation provides a clearer visualization of the data, since it is a continuous function of its variables, it has very reduced noise, its precision does not depend on the size of the bins of the distribution's histogram and it has a better
defined cutoff \cite{newmanPowerlaw}. Here, $s$ is the amount of spikes
in an avalanche and $t$ is the amount of time windows during which the avalanche took place.
A given data set with probability distribution function $P(s)= Bs^{-\a}$ and cutoff $Z$ ($B$ is constant)
corresponds
to a cumulative distribution
\begin{equation}
\label{eq:cumulativeFit}
P(s>S) = A + B's^{-\a'}\ \textnormal{,}
\end{equation}
such that $\a=\a'+1$, $B'=B/\a'$ and $A=-BZ^{-\a'}/\a'$.

All results refer to square lattices of linear size
$L$ with free boundary conditions and nearest neighbor couplings.
The initial conditions for all neurons are the fixed point $(x^*,y^*,z^*)$ for a given set of parameters.
The initial conditions $\left(I^{syn}_{ij}(0),h_{ij}(0)\right)$ for the synapses are set to zero. 

\begin{figure}[t!]
	\includegraphics[width=60mm]{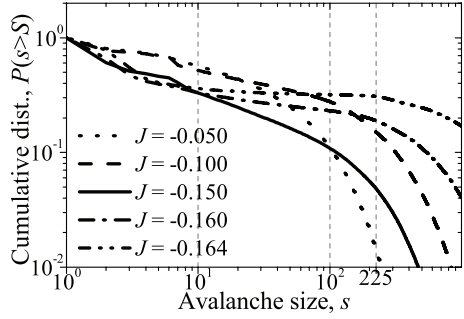}
	\caption{\label{fg:reg1DifJSizes}Avalanche size cumulative distributions for neurons in regime I, $p=0.3$,
	$L=15$ and different values of $J$. There is a clear change in the shapes of the curves as $J$ is increasing.}
\end{figure}

Some dynamical features of neurons and synapses have revealed themselves very important for the occurrence of
critical avalanches, in special the size of the refractory period and the synapses characteristic times. 
If the synapse takes longer to excite the neighbor than the duration of the refractory period of the
presynaptic neuron, then the wave of activity propagates forward and backward in the network, producing
self-sustainend activity in the form of spiral waves. This reasoning guided us in choosing the following
neuron and synapse sets of parameters.

For each simulation, all neurons and synapses have the same parameters.
We examine three different excitable regimes: (I) $x_R=-0.7$, $\l=0.008$ -- neurons can be excited either by positive and by negative inputs, which generates
rebound spikes; (II) $x_R=-0.9$, $\l=0.01$; and (III) $x_R=-0.9$, $\l=0.1$ -- both regimes II and III can be excited only by
positive inputs, but they have different refractory periods. The remaining parameters of the neurons are always $K=0.6$, $T=0.35$ and $\d=0.001$.

The synapses are fast (time constants $\t_1=\t_2=2$ time steps), whereas the spike half-duration
takes $\approx 6$ time steps \cite{modeloKTz2001}). If we use a typical value of $1$ ms for the half-duration, we can set the
time scale (1 time step or 1 ts = 1/6 ms) and get $\t_1 \approx 0.33$ ms which is also typical for fast synapses \cite{synapticModelSchutter}.
We studied inhibitory ($J<0$) and excitatory ($J>0$) synapses for regime I, and excitatory synapses for regimes II and III.

The network is always stimulated in a randomly chosen site. To separate the time scales, we impose that each stimulus
happens only after the end of the previous avalanche. The stimulus takes
place during $1$ ts (a delta stimulus) with intensity $I_{ext}$ sufficient to produce a spike. We use $I_{ext}=0.1$ for
regime I and $I_{ext}=0.4$ for regimes II and III. The simulation is divided in time windows of $20$ ts each. 
These windows are used to count the spikes in the avalanches, just like in the
experimental protocol \cite{violaSub,ribeiroCopelli}.

\begin{figure}[t!]
	\includegraphics[width=85mm]{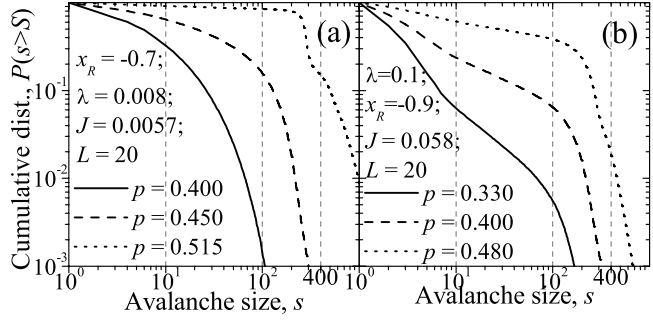}
	\caption{\label{fg:reg13Sizes}Avalanche size cumulative distributions for neurons in regime I (a) and regime III (b)
	for excitatory synapses, $L=20$ and for different excitation probability $p$. Parameters are $J=0.0057$ (a) and 
	$J=0.058$ (b).}
\end{figure}

Avalanche size cumulative distributions, $P(s>S)\sim s^{-\a'}$, for $L=15$, $L=20$ and $L=30$ are shown in Fig. \ref{fg:reg1L20L30SizeTime} (a) whereas the
duration cumulative distributions, $P(t>T)\sim s^{-\t'}$, are in Fig. \ref{fg:reg1L20L30SizeTime} (b), both for neurons in regime I with
excitatory by rebound synapses.
Fitting Eq. \ref{eq:cumulativeFit} to the curves in Fig. \ref{fg:reg1L20L30SizeTime}, gives $\a=\a'+1=1.35$, $\t=\t'+1=1.50$ and a spatial cutoff $Z_s=L^\g$, with $\g=2.46\pm0.02$.
Since the avalanches propagate like spiral waves, we expect $\g>2$, as the same neuron may participate
more than once in a given avalanche.

\begin{figure}[b!]
	\includegraphics[width=85mm]{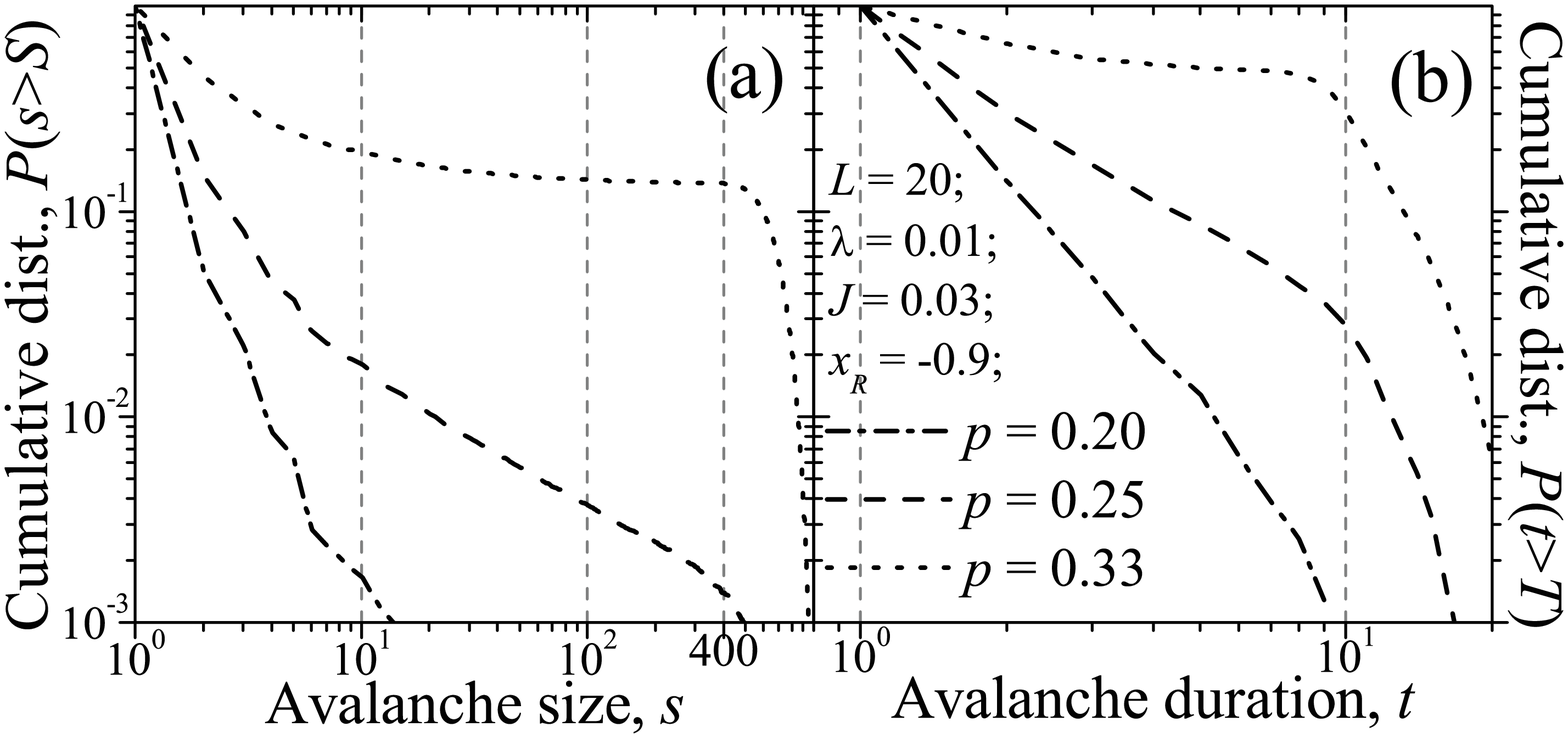}
	\caption{\label{fg:reg2Exc}Avalanche size cumulative distributions (a) and duration cumulative distributions (b)
	for neurons in regime II for excitatory synapses ($J=0.03$) and different excitation probabilities with $L=20$.}
\end{figure}

Fig. \ref{fg:reg1DifJSizes} suggests a critial region in the $p\times J\times R$ space (which are related by Eq. \ref{eq:excProb}),
since for a given $p$, we observe three regimes.
In this case, $p=0.3$ and the power law avalanches regime (for $J=J_c=-0.15$) sits between two regimes: one with predominance of small avalanches
($J>J_c$) and the other with preeminence of big avalanches ($J<J_c$).

Figs. \ref{fg:reg13Sizes} and \ref{fg:reg2Exc} show the cumulative distribution of the avalanche sizes for regime I and III and of the
avalanche sizes and durations for regime II. None of the curves may be fit by Eq. \ref{eq:cumulativeFit}, so there is no critical behavior.
In fact, these results agree with other authors who have shown that in purely excitatory networks, the cutoff is much smaller than
the network size \cite{beggsPlenz2003}.

\begin{figure}[t!]
	\begin{center}
	\includegraphics[width=60mm]{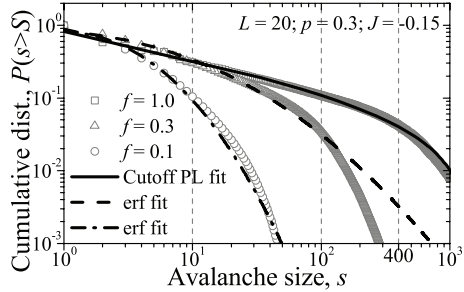}
	\end{center}
	\caption{\label{fg:reg1Sub}Avalanche size cumulative distribution for a complete sampling (------) and subsamplings of
	fractions $f=0.3$ ($---$) and $f=0.1$ ($-\cdot-$) of neurons in regime I. Symbols are simulation data. Subsamplings are fitted by
	error-function curves whereas the complete sampling is by Eq. \ref{eq:cumulativeFit}.}
\end{figure}

The subsampling effect is shown in Fig. \ref{fg:reg1Sub} for regime I with excitatory by rebound synapses for different sampling
fractions, $f$. For $f\leq0.1$, the avalanche size cumulative distributions, $P(s>S)$, match an error-function fit,
which corresponds to a lognormal distribution, $P(s)$, found in
cellular automata models and experiments \cite{ribeiroCopelli}.

Since rebound spikes are delayed compared to excitatory spikes,
we could only produce power law avalanches with excitatory by rebound synapses (Figs. \ref{fg:reg1L20L30SizeTime} and \ref{fg:reg1DifJSizes}).
Otherwise, the avalanches are much smaller than the network size (Figs. \ref{fg:reg13Sizes} and \ref{fg:reg2Exc}).
We also showed that synaptic noise is a new way of generating critical avalanches
(one would expect it for the same reason that disorder may change a first order phase transition into a second order one
\cite{disorderPhase}). Therefore, criticality
may be a product of the stochasticity in synaptic interactions, as the noise dissipates the activity just like
the inhibitory synapses do in excitatory-inhibitory balanced models \cite{vertesCNS}.

Our map-based  model presents an out of equilibrium phase transition which we conjecture, following Bonachela \textit{et. al} \cite{bonachela2},
to pertain to the dynamical percolation universality class. 
Our next efforts will be to unveil the critical region in the $p\times J$ plane, to study different topologies and
heterogeneous networks (mixing excitatory with inhibitory directed synapses). We may also add an extra dynamical rule in the
noise amplitude $R$ in order to self-adjust it towards the critical region.
Due to the complexity of our model and the
correspondence with more biological features, we hope to provide clues on what type of neurons and what type of synapses
could show criticality in the brain. Then, one can check whether these characteristics are present in experimental situations, like our prediction that 
critical avalanches could be observed in excitatory by rebound networks with fast synapses if
neurons produce rebound spikes.

We thank M. Copelli, A. Roque da Silva, D. Arruda and
V. Priesemann for discussions.

\end{document}